\documentclass[12pt]{article}
\usepackage{graphicx}
\usepackage{epsfig}

\begin{document}

\newcommand{\beq}{\begin{equation}}
\newcommand{\eeq}{\end{equation}}
\newcommand{\bea}{\begin{eqnarray}}
\newcommand{\eea}{\end{eqnarray}}
\newcommand{\eps}{\varepsilon}
\newcommand{\Fs}{\mbox{\scriptsize F}}
\newcommand{\lsim}{\stackrel{\scriptstyle <}{\phantom{}_{\sim}}}
\newcommand{\gsim}{\stackrel{\scriptstyle >}{\phantom{}_{\sim}}}

{\bf Upper edge of the neutron star inner crust: the drip point
and around}

\vskip 0.5 cm \noindent M.~Baldo$^{1}$, E.E.~Saperstein$^{2}$ and
S.V.~Tolokonnikov$^{2}$

\vskip 0.5 cm

\noindent $^1$INFN, Sezione di Catania, 64 Via S.-Sofia, I-95123
Catania,
Italy \\
$^{2}$ Kurchatov Institute, 123182, Moscow, Russia

\vskip 0.5 cm
\begin{abstract}

A semi-microscopic self-consistent quantum  approach developed recently to
describe the inner crust structure of neutron stars within the  Wigner-Seitz
method and the explicit inclusion of neutron and proton pairing correlations is
used for finding the neutron drip point which separates the outer and inner
crusts. The equilibrium configurations of the crust are examined in vicinity of
the drip point and in the upper part of the inner crust, for the density region
corresponding to average Fermi momenta $k_{\rm F}{=}0.2 \div 0.5\;$fm$^{-1}$.

\end{abstract}

%\PACS{21.60.-n; 21.65.+f; 26.60.+c}

\section{Introduction}

The crust of a neutron star consists of matter with subnuclear densities, $\rho
\le 0.5 \rho_0 $, where $\rho_0 \simeq 0.17 \,\mbox{\rm fm}^{-3}$ is the
nuclear saturation density. The outer part of the crust is a crystal system
consisting of mainly spherically symmetrical nuclei immersed in a  virtually
uniform sea of ultra-relativistic electrons which makes the system
electro-neutral. At some critical density value, the so-called neutron drip
point $\rho_{\rm d}$, the neutron chemical potential becomes positive. As a
consequence, in the inner part of the crust a portion of neutrons begins to
become delocalized and forms a superfluid neutron liquid surrounding the
nuclear-like clusters. As it is well known, the neutron superfluidity is
responsible for many properties of neutron stars \cite{Pet}.

The crust contains only about 1\% of the neutron star mass, its
thickness being only 10\% of the neutron star radius. However, the
understanding  of the crust structure is of primary importance for
explaining several important observational phenomena in neutron
stars. Sudden changes of the pulsar rotation periods, the
so-called glitches, should be mentioned here in the first place.
The neutron superfluidity of the crust is also very important for
the cooling process of the neutron star \cite{Pet}.

The drip point was originally calculated by Baym, Pethick, and Sutherland (BPS)
\cite{BPS} who developed a thermodynamic approach to describe the outer crust
using a phenomenological nuclear equation of states based on an extrapolation
of the nuclear mass data. Their prediction for the drip point was $\rho_{\rm
d}{=}4.3\cdot 10^{11}$g/cm$^3$ which corresponds to the average Fermi momentum
$k_{\rm F}^{\rm d }{=}0.1977 \;$fm$^{-1}$. \footnote{The average Fermi momentum
is defined as $k_{\rm F}{=}(3\pi^2\rho)^{1/3}$, corresponding to homogeneous
neutron matter at the same average density. We prefer to deal with this
variable instead of $\rho$ because, through the inner crust, it changes much
more slowly than the density itself, which makes smoother the graphical
presentation of the results.} A little later, Negele and Vautherin (NV)
\cite{NV} developed a completely quantum approach to describe the inner crust
of a neutron star based on the energy density functional method in combination
with the spherical Wigner-Seitz (WS) approximation.
 Within this method, the crystal matter under
consideration is approximated with the set of independent
spherical cells, the neutron single-particle wave functions
obeying some boundary condition.  This method  was applied in
\cite{NV} to the self-consistent description of the structure of
the inner crust in a wide density region. NV confirmed also the
BPS prediction of the drip point value, the drip instability in
their study occurring for the nucleus $^{118}_{36}$Kr. The first
point just above the drip point ($k_{\rm F}{=}0.202 \;$fm$^{-1}$)
considered in \cite{NV} corresponds  to an equilibrium
configuration with $Z{=}40$.

Recently  S. B. Ruster et al. \cite{Rust} carried out a systematic analysis of
the outer crust structure in the vicinity of the drip point within the
thermodynamic BPS method, but with a set of different models for the nuclear
equation of state (various kinds of Skyrme forces, modern versions of the
droplet model, relativistic nuclear mean field theory and so on). All the
predictions for the neutron drip point turned out to be rather close to the BPS
one, $\rho_{\rm d} \simeq 4 \cdot 10^{11}$g/cm$^3$. As to the drip $Z$ values,
$Z{=}34\div 38$, they are quite close to the NV ones.

In this paper, we try to study the drip point region within a semi-microscopic
self-consistent quantum approach which was developed recently
\cite{crust1,crust2,crust3,crust4} to describe the inner crust. This approach
is generically close to the NV one but  takes into account in a self-consistent
way the effect of the neutron and proton pairing correlations. Within the NV
approach, for a fixed average nuclear density $\rho$, the energy functional of
the system is minimized for the spherical WS cell of the radius $R_c$. Each
cell contains \beq A=(4\pi/3)R_c^3\rho
 \label{Anumb}
\eeq nucleons. This number is shared between $Z$ protons  and
$N=A-Z$ neutrons. As far as the neutron star matter is uncharged,
the cell should contain $Z$ electrons.  In addition, the
$\beta$-stability condition, \beq \mu_e= \mu_n-\mu_p,
 \label{beta}
\eeq
 has to be fulfilled, where $\mu_e$, $\mu_n$ and
$\mu_p$ are the chemical potentials of electrons, neutrons and protons,
respectively. They are defined as to include the corresponding bare masses. In
the NV calculation, pairing effects were not taken into account because their
contribution to the total energy of the system under consideration is quite
small. Only 30 years after the classical paper \cite{NV}, in 2004, there
appeared a self-consistent calculation of the inner crust structure with the
pairing effects taken into consideration \cite{MMM}. It was carried out within
the Hartree-Fock method with Skyrme forces plus the BCS approximation employing
some schematic effective pairing forces. In \cite{MMM}, the main attention was
paid to the comparison of the neutron pairing gap in the  crust with that for
the homogeneous neutron matter. However, the influence of the pairing effects
on the equilibrium configuration ($Z,R_c$) of the inner crust at different
densities was not analyzed.

This effect was demonstrated in \cite{crust1,crust2} and analyzed
systematically in \cite{crust3,crust4}. In the approach formulated
in \cite{crust3} the generalized energy functional method
\cite{STF,Fay} is used, which incorporates the pairing effects
into the original Kohn-Sham \cite{KS} method. The interaction part
of the generalized energy functional depends, on equal footing, on
the normal densities $\rho_n , \rho_p$, and the anomalous ones,
$\nu_n, \nu_p$, as well. The ansatz of \cite{crust3} for the
complete energy functional is  a smooth matching of the
phenomenological and the microscopic functionals at the nuclear
cluster surface. For the normal component of the first one, we use
the nuclear effective functional DF3 by S. A. Fayans et al.
\cite{Fay}. The anomalous term of the nuclear functional of
\cite{Fay} is modified a little in \cite{crust3} to the form which
is more adequate for describing the inner crust structure. The
microscopic part of the energy functional which describes the
neutron matter surrounding the nuclear cluster is calculated
within the Brueckner approach with the Argonne force v$_{18}$
\cite{Bal}. The systematic analysis performed for a wide density
region showed that, as a rule, the equilibrium configuration
($Z,R_c$) changes noticeably due to pairing. The effect grows as
the density increases, and for the equilibrium $Z$ values at
$k_{\rm F}\simeq 1\;$fm$^{-1}$ reaches a factor of two. So strong
pairing effect can be explained on the basis of the
$\beta$-equilibrium condition (\ref{beta}). In fact, the
variations due to pairing of the chemical potentials, $\mu_n$ and
$\mu_p$, are, as a rule, much stronger than those of the total
binding energy.

In \cite{crust3,crust4}, the intermediate and bottom parts of the inner crust
was examined in the density interval corresponding to the average Fermi momenta
$k_{\rm F}{=}0.6 \div 1.2\;$fm$^{-1}$. In this paper, we use the same method
for finding the drip point $\rho_{\rm d}$ and for investigating the ground
state configurations of the crust in vicinity of this density value. To make
description of the inner crust more complete, we carried out also systematic
calculations for the interval $k_{\rm F}{=}0.2 \div 0.5\;$fm$^{-1}$. In
connection with validity of the WS approximation, it is worth to mention that
recently a consistent band theory was developed for the inner star crust
\cite{Cham1,Cham2,Cham3}. In \cite{Cham4}, this more fundamental method is
compared with the WS approximation for the upper region of the inner crust. It
is concluded that the WS approximation is well suited for describing the ground
state properties of the system under consideration, whereas the description of
some dynamical aspects needs the consistent band theory. As far as we limit
ourselves to the ground state properties, we can rely on the validity of the WS
method for the density interval under consideration.

\section{On the method}
The method we developed is described in detail in Refs.
\cite{crust3,compar,P123}. Here we specify only some details to
explain the version we use in this paper. The ansatz of
\cite{crust3} for the complete energy functional is  a smooth
matching of the phenomenological and the microscopic functionals
at the cluster surface:

\beq {\cal E}(\rho_{\tau}({\bf r}),\nu_{\tau}({\bf r})) = {\cal E}^{\rm
ph}(\rho_{\tau}({\bf r}),\nu_{\tau}({\bf r})) F_m( r)+ {\cal E}^{\rm
mi}(\rho_{\tau}({\bf r}),\nu_{\tau}({\bf r}))(1 - F_m(r)), \label{tot} \eeq
where the matching function $F_m(r)$ is a two-parameter Fermi function: \beq
F_m(r)=(1+\exp((r-R_m)/d_m))^{-1}. \label{match} \eeq \noindent Eq. (\ref{tot})
is applied  both to the normal and to the anomalous components of the energy
functional. The diffuseness parameter was taken equal to $d_m{=}0.3\;$fm for
any value of the average baryon density of the inner crust and for any
configuration ($Z,R_c$). The matching radius $R_m$ is chosen anew in any new
case in such a way that the equality \beq \rho_p(R_m){=}0.1 \rho_p(0)
\label{Rmatch} \eeq  holds.

We use the nuclear DF3 functional for the phenomenological
component of (\ref{tot}). For the microscopic part of the normal
component of the total energy functional (\ref{tot}) we follow
refs. \cite{crust3} and take the equation of states of neutron
matter calculated in \cite{Bal} with the Argonne v$_{18}$
potential on the basis of Brueckner theory, taking into account a
small admixture of 3-body force. Its explicit form can be found in
the cited articles.

 The anomalous part of the energy functional used in
\cite{crust3} has the form: \beq {\cal E}_{\rm an} =\frac {1}{2}
\sum_{\tau} {\cal V}_{{\rm an},\tau}^{\rm eff}(r) |\nu_{\tau}({\bf
r})|^2, \label{an} \eeq where ${\cal V}_{{\rm an},\tau}^{\rm eff}$
($\tau{=}n,p$) is the density dependent effective pairing
interaction.

 The matching relation (\ref{tot}) for the anomalous part of the
 energy functional leads to the analogous relation for the effective
pairing interaction:
 \beq {\cal V}_{\rm an}^{\rm eff}(r)= {\cal V}_{\rm eff}^{\rm ph}(\rho(r)) F_m( r) +
 {\cal V}_{\rm eff}^{\rm mi}(\rho(r))(1-F_m( r)).
\label{mEPI} \eeq The isotopic index $\tau$ is for brevity omitted. The
phenomenological effective pairing interaction ${\cal V}_{\rm eff}^{\rm ph}$
in \cite{crust3} has a density dependent coordinate delta-function form, just
as in \cite{Fay}. The explicit form of the density dependence \cite{Fay} was
modified a little in \cite{crust3} in accordance with (\ref{mEPI}). As to the
microscopic effective pairing interaction, it was calculated in \cite{crust3}
within the BCS approximation with the same Argonne force v$_{18}$ as the normal
part of the energy functional. In this approach, the set of the gap values
$\Delta^{\rm BCS}(k_{\Fs})$  for all the Fermi momenta under consideration is
the only input for finding the effective pairing interaction. In \cite{P123}, a
more realistic model for neutron pairing was used to take into account the
many-body corrections to the BCS approximation which, as it is commonly known,
suppress the neutron gap significantly. The density and momentum independent
suppression factor was introduced, \beq \Delta_n(k,k_{\Fs}) = f_{\rm m-b}
\Delta_n^{\rm BCS}(k,k_{\Fs}), \label{fac} \eeq to describe the many-body
 effects approximately.
Two versions of this model were considered, with $f_{\rm m-b}{=}1/2$ (P2 model)
or $f_{\rm m-b}{=}1/3$ (P3 model). In this notation, the BCS approximation has
been named as the P1 model. In this paper, we limit ourselves with the
intermediate case of the P2 model.

One more remark should be made before presenting the results. As
it is known \cite{NV}, the boundary conditions for the neutron
single-particle wave functions inherent to the  WS method are more
or less arbitrary provided they guarantee the orthogonality and
completeness of the function set.  NV used the following one:
 \beq
 R_{nlj}(r=R_c)=0
 \label{bco}
\eeq for odd $l$, and \beq   R'_{nlj}(r=R_c)=0, \label{bce} \eeq
 for even ones.
 In \cite{compar} two kinds of these conditions were examined which
{\it a priory} look similar. The NV boundary condition was denoted
as BC1. An alternative kind of the boundary conditions (BC2) was
considered in \cite{compar} also, when Eq.~(\ref{bco}) is valid
for even $l$ whereas Eq.~(\ref{bce}), for odd ones. The analysis
of \cite{compar} for the BCS approximation (the P1 model) and the
similar one of \cite{P123} for the P2 and P3 models have shown
that predictions on the basis of BC1 {\it versus} BC2 are in
general different. This results in uncertainties inherent to the
WS method itself which grow at increasing $k_{\Fs}$. But in the
upper part of the inner crust they turned out to be negligible
\cite{compar}. Therefore in this paper we limit ourselves to the
BC1 version of boundary conditions.

\section{The drip point}

The approach of \cite{crust3} based on the WS method was developed initially
for describing the inner crust. In this case, part of the neutrons becomes
delocalized, therefore the neutron number $N$ per WS cell can be fractional.
The same is true for the total nucleon number per cell found from
(\ref{Anumb}). In this case, the $\beta$-stability relation (\ref{beta}) always
has a solution as far as two chemical potentials in this equation, $\mu_e$ and
$\mu_n$, are continuous regular functions of the cell radius $R_c$.

The WS method could be used also to the outer crust, $\rho <\rho_{\rm d}$, with
some small modifications. Now, all neutrons in the system are bound inside the
nuclei, therefore $N$ should be integer. Then, the chemical potentials $\mu_p$
and $\mu_n$ are discrete functions of $(N,Z)$ and both don't depend practically
on $R_c$. Therefore, for an arbitrary density $\rho$, Eq.~(\ref{beta}) has no
exact solution.  However, it is not necessary in this case. Instead of
Eq.~(\ref{beta}), the $\beta$-stability condition for the outer crust looks  as
a set of inequalities \beq E_0(N,Z) - E_0(N-1,Z+1) \le \mu_e \le E_0(N+1,Z-1) -
E_0(N,Z), \label{beta1} \eeq where $E_0(N,Z)$ is the ground state energy of the
corresponding nucleus. $E_0$ differs from the binding energy $E_{\rm B}$ by the
inclusion of the bare neutron and proton masses. In the limit of large $N,Z$
values, when one can neglect the difference of $\mu_n,\mu_p$ in neighboring
nuclei, the double inequality (\ref{beta1}) yields the equality (\ref{beta}).
Estimations show that in the bottom part of the outer crust, in the vicinity of
the drip point, one can approximately use the same version of the WS method as
that for the inner crust. In particular, one can suppose $N$ (and $A$) to be
fractional, which leads to the possibility of solving Eq.~(\ref{beta}) exactly.
In this Section, we do use such an approximate scheme.

It is worth to mention that in the outer crust  the
phenomenological component of the energy functional (\ref{tot})
plays the main role. Indeed, as far as all neutron are now bound,
the microscopic term of (\ref{tot}) is switched on only in the
region where the neutron density falls. However, we can not {\it a
priory} neglect this contribution. In the physical situation of
the outer crust the complete functional (\ref{tot}) is practically
equivalent to a modification of the surface components of the
effective interaction appearing in the phenomenological functional
of \cite{Fay}, but it is well known that predictions of the
self-consistent calculations for atomic nuclei are highly
sensitive to details of the effective interaction at the surface.
Therefore, for finding the drip point we carried out two sets of
calculations. In the first one we used only the phenomenological
nuclear DF3 functional, in the second one, the complete
semi-microscopic functional of (\ref{tot}).

Let us first consider in more detail the version with the pure
phenomenological DF3 functional. In Fig. 1, the binding energy per
nucleon $E_{\rm B}$ is displayed as a function on $Z$ for
different values of $k_{\rm F}$ in vicinity of the critical point
by BPS. As one can see, the minimum of the function $E_{\rm B}(Z)$
is shifted from the NV value of $Z{=}36\div 40$ (or $Z{=}34\div
38$ of \cite{Rust}) to $Z=52\div54$. The corresponding curves for
the $\mu_n(Z)$ dependence are displayed in Fig. 2. At each value
of $k_{\rm F}$, the star indicates the point corresponding to the
minimum of $E_{\rm B}(Z)$, i.e. to the equilibrium configuration.
As one can see, the drip point corresponds to the critical value
of the Fermi momentum which is a little above $k_{\rm F}{=}0.18
\;$fm$^{-1}$. A simple interpolation results in the value of
$k_{\rm F}^{\rm d } {=} 0.181 \;$fm$^{-1}$. The latter corresponds
to the density $\rho_{\rm d}{=}3.30\cdot 10^{11}$g/cm$^3$. Thus,
the difference from the BPS prediction (and that of \cite{Rust})
for the drip point position in this case is not negligible. But
the most essential deviation occurs for the ground state
configuration ($Z,R_c$) at the drip point and around. The main
characteristics of the equilibrium WS cell for the case under
consideration are presented in Table 1. As one can see, the drip
point predicted with the DF3 functional corresponds to the drip
nucleus with ($Z,N$) values in vicinity of the magic numbers
$Z{=}50$ and $N{=}126$. Fig. 2 shows that there is another "drip
region" with $Z\simeq 40$ (corresponding neutron number values are
in vicinity of the magic number $N{=}82$). However, as it is seen
from Fig.1, this region corresponds to higher values of the
binding energy.

\begin{table}
\caption{  Characteristics of the equilibrium WS cell in vicinity
of the drip point for the case of the phenomenological DF3
functional \cite{Fay}.}
\bigskip

\begin{tabular}{|c|c|c|c|c|c|}
\hline
 $ k_{\Fs}$, fm$^{-1}$ & $Z$  &$N$ &   $R_c$, fm & $E_{\rm B}$, MeV &$\mu_n$,
 MeV
\\
\hline

  0.160 & 54 & 125 & 67.61 &  -1.820  & -0.957  \\
  0.170 & 54 & 127 & 63.82 &  -1.544  & -0.355  \\
  0.175 & 52 & 126 & 61.64 &  -1.408  & -0.436   \\
  0.180 & 52 & 126 & 59.98 &  -1.275  & -0.062   \\
  0.185 & 52 & 128 & 58.61 &  -1.145  &  0.487   \\

\hline
\end{tabular}
\end{table}

Then we repeated the calculations with the complete functional (\ref{tot})
including both the phenomenological DF3 and microscopic components. The results
are given in Table 2. In this case, the drip point is shifted a little to
$k_{\rm F}^{\rm d } = 0.194 \;$fm$^{-1}$ which corresponds to the density
$\rho_{\rm d}{=}4.06\cdot 10^{11}$g/cm$^3$. It is quite close to the value by
BPS and it is within the interval given in \cite{Rust}. As to the $Z$ and $N$
values, they are practically the same as in Table 1.

\begin{table}
\caption{ Characteristics of the equilibrium WS cell in vicinity
of the drip point for the case of the semi-microscopic functional
\cite{crust3}.}
\bigskip

\begin{tabular}{|c|c|c|c|c|c|}
\hline
 $ k_{\Fs}$, fm$^{-1}$ & $Z$  &$N$&   $R_c$, fm & $E_{\rm B}$, MeV &$\mu_n$,
 MeV
\\
\hline

  0.180 & 52 & 128  & 60.14 &  -1.398  & -0.868   \\
  0.190 & 52 & 126  & 57.85 &  -1.150  & -0.552   \\
  0.195 & 52 & 126  & 57.36 &  -1.043  & 0.123   \\
  0.200 & 52 & 130  & 57.19 &  -0.950  & 0.194   \\
\hline
\end{tabular}
\end{table}

Thus, our results obtained within the self-consistent quantum approach, with
both the phenomenological DF3 functional  or the semi-microscopic one
\cite{crust3}, agree approximately with predictions of \cite{Rust} for the drip
point position but deviate significantly for the drip $Z$ values. Hardly the
latter originates from the difference between the thermodynamic BPS approach
used in \cite{Rust} and the completely quantum method (\'a l\'a the NV one)
used in this paper. Indeed, the drip $Z$ NV values  found within, in general,
similar quantum approach are very close to those in \cite{Rust}. The main
origin of the difference is related, evidently, to the peculiarities  of the
DF3 functional. It should be stressed that this functional describes with high
accuracy a lot of spherical nuclei, including long isotopic chains \cite{Fay}.
It is worth to mention also that recently Y.Yu and A.Bulgac \cite{Bul1}
examined binding energies of a number of isotope and isotone chains (with the
total number of nuclei exceeding 200) within the so-called SLDA (Superfluid
LDA) method developed previously by A. Bulgac\cite{Bul2}.  For the normal
component of the energy density functional they used alternatively or the
popular version of the Skyrme force, SLy4, or the density functional  FaNDF$^0$
by S.A. Fayans \cite{Fay1}. \footnote{The DF3 functional of \cite{Fay} is an
advanced version of FaNDF$^0$, but the difference is quite small.} For both
versions, the authors of \cite{Bul1} obtained very good description of the
data, but the accuracy of the run with the FaNDF$^0$  force turns out to be
higher. That is why we think that our results for the drip point region for
both sets of calculations are closer to the truth. Remind that we used the DF3
functional in the first one and the semi-microscopic functional of
\cite{crust3,compar,P123} in the second one. The latter functional contains the
same DF3 functional as the phenomenological ingredient of (\ref{tot}).

\section{The upper part of the inner crust}

Let us go now to densities above the drip point corresponding to
the upper part of the inner crust. In this region, we have
considered previously \cite{compar} only the case of $k_{\rm F}
{=} 0.2 \;$fm$^{-1}$  which is very close to the drip point. Now,
we analyze systematically the region of $k_{\rm F} {=} 0.2 \div
0.5 \;$fm$^{-1}$. As it was mentioned above, we limit ourselves
with the P2 model of\cite{P123} and with BC1 boundary condition of
the WS method (i.e. that of \cite{NV}).
 Results are
presented in Figs. 3 - 6 and in Tables 3 - 6. In Table 3, the
equilibrium characteristics of the inner crust within the WS
method are given. The ratio $x{=}Z/A$ indicates the average
concentration of protons for a density under consideration. For
the sake of comparison, the corresponding NV values  are also
presented. It should be noted that calculations in \cite{NV} were
performed at slightly different values of the density. We took the
results of \cite{NV} for the closest values of the average Fermi
momentum $k_{\rm F}{=}$0.202, 0.296, 0.361 and 0.480 fm$^{-1}$,
correspondingly. The difference between predictions of our
approach and the NV ones for the $Z$ and $A$ values turned out to
be of the same order of magnitude as in the intermediate and
bottom parts of the crust ($k_{\rm F}\ge 0.6\;$fm$^{-1}$)
investigated in \cite{crust4,P123}. In the latter case, we
explained this difference mainly with the neutron pairing effects.
For the upper layers of the inner crust their influence  and all
the role  of the neutron surroundings  is smaller, therefore the
main reason of the difference is, evidently, the difference
between the DF3 functional and the NV one, when the latter is
applied  to ordinary atomic nuclei.

\begin{table}
\caption{  Characteristics of the equilibrium WS cell in the upper
part of the inner crust.}
\bigskip

\begin{tabular}{|c|c|c|c|c|c|c|c|c|}
\hline
 $ k_{\Fs}$, fm$^{-1}$ & $Z$  & $Z$ \cite{NV}&  $A$ &  $A$ \cite{NV} &
   $R_c$, fm &  $R_c$, fm \cite{NV}&
 $x$ &$x$ \cite{NV} \\

\hline

  0.2 & 52 & 40&  212 & 180 &  57.2 & 53.6 & 0.245  & 0.222   \\
  0.3 & 54 & 40&  562 & 320 &  52.8 & 44.3 & 0.096  & 0.125   \\
  0.4 & 50 & 40&  830 & 500 &  45.1 & 42.2 & 0.060  & 0.080   \\
  0.5 & 46 & 40& 1020 & 950 &  38.6 & 39.3 & 0.045  & 0.042   \\

\hline
\end{tabular}
\end{table}

\begin{table}
\caption{  The ground state energy characteristics of the matter
in the upper part of the inner crust. }
\bigskip

\begin{tabular}{|c|c|c|c|c|}
\hline
 $ k_{\Fs}$, fm$^{-1}$ & $Z$  &  $A$  &
 $E_{\rm B}$, MeV &$\mu_n$, MeV\\

\hline

  0.2 & 52 &   212    & -0.950  & 0.194   \\
  0.3 & 54 &   562    &  0.211  & 1.018   \\
  0.4 & 50 &   830    &  0.950  & 1.804   \\
  0.5 & 46 &  1020    &  1.625  & 2.643   \\

\hline
\end{tabular}
\end{table}

Table 4 contains the main energy characteristics of the inner crust in the
region under consideration.  Some average characteristics of nuclear clusters
in the center of the WS cell are presented in Table 5. Here $R_m$ is the
matching radius defined with Eq.~(\ref{Rmatch}). To avoid influence of quantum
oscillations, the quantity $\rho_p(0)$ in this relation was replaced by the
average proton density over the interval $r<3\;$fm. We include into the cluster
all nucleons inside the sphere with the radius $R_m$. \footnote{It should be
stressed that the dividing of the WS cell into the cluster and surrounding
neutron matter is used here just for an illustration, and the calculation
procedure outlined above doesn't involve any explicit separation procedure.} It
contains practically all Z protons and \beq N_{cl}= \int_{r<R_m}d^3r \rho_n(r).
\label{Ncl}\eeq neutrons. The total particle number in the cluster is
$A_{cl}=Z+N_{cl}$. The ratio $x_{cl}{=}Z/A_{cl}$ is the average proton
concentration inside the cluster and $y_{cl}{=}A_{cl}/A$, the relative nucleon
number inside the cluster. At last, the neutron and proton cluster radii $R_n$
and $R_p$ are defined as the points of the maximum gradient of the
corresponding densities. These densities are displayed in Fig.4 (neutrons) and
Fig. 5 (protons). For comparison, the similar densities for ordinary atomic
nuclei with the same $Z$ and $A{=}A_{cl}$ were also calculated and displayed
(with the dotted lines). One can see that, in the density region under
consideration, the nuclear-type clusters in the center of the WS cell are quite
similar to their nuclear counterparts. All such nuclei are bound with the
exception of the one for the case of $ k_{\Fs}{=}0.5\;$fm$^{-1}$,
$^{46}_{166}$Pd, which occurred a little above the corresponding neutron drip
line (the neutron chemical potential is $\mu_n \simeq 0.1\;$MeV). In the
self-consistent calculation procedure for such unstable nucleus we put it in an
external box. As far as the contribution of the positive energy states to the
total neutron density for such a small value of $\mu_n$ is tiny, it could be
neglected for a qualitative comparison. At higher values of $ k_{\Fs} \ge
0.6\;$fm$^{-1}$ considered in \cite{crust4,P123} all the nuclear counterparts
of the nuclear clusters are already unbound significantly, which makes a
similar comparison impossible. The main difference of the density distribution
in the cluster from that in the nuclear counterpart is  the bigger values of
the radius and of the diffuseness parameter.  This is true both for neutrons
and for protons, for all  $ k_{\Fs}$ values under consideration, the effect
being lager for higher density values. At high values of $k_{\Fs}$ analyzed in
\cite{crust4,P123} the quasi-nuclei in the centers of the WS cells have nothing
in common with ordinary nuclei \cite{crust4}.

Let us go now to the analysis of the neutron gap  function $\Delta_n(r)$,
which, as it was mentioned in the Introduction, is of primary importance for
neutron star observable properties. The gap functions for different values of
$k_{\rm F}$ are displayed in Fig. 6. One can see that, although at each value
of $k_{\rm F}$ there is a non-regular behavior of $\Delta_n(r)$ at the central
cluster surface, the absolute value of the gap inside the cluster is governed
by that in the asymptotic region and it is essentially different for different
$k_{\rm F}$. This feature of the $\Delta_n(r)$ function contradicts to a naive
LDA picture. Within such a naive LDA, dealing with the combined
phenomenological-microscopical energy functional of the (\ref{tot}) form, one
could expect the gap function inside the cluster, where only the
phenomenological component of the energy functional survives, close to that in
ordinary atomic nuclei which is typically of the order of 1 MeV. Indeed, as we
see in Fig. 4, the neutron density distribution $\rho_n(r)$ inside the cluster
is very close to that in the nuclear counterpart. The drastically different
pattern in Fig. 6 demonstrates a strong inside-outside interplay in the gap
equation, the so-called proximity effect.

Table 6 collects some important characteristics of the gap for all the density
values under consideration. The notation of \cite{P123} is used. For sake of
completeness, we repeat it shortly. The asymptotic value of the Fermi momentum
$k_{\rm F}^{\rm as}$ is calculated in terms of the asymptotic value of the
density $\rho(r)$ averaged over the interval $R_c{-}b < r < R_c $, $b{=}2\;$fm.
The analogous average is calculated for the asymptotic gap value $\Delta_{\rm
as}$. The central gap value $\Delta(0)$ is determined as the average of
$\Delta(r)$ over the interval $0<r<3\;$fm. The Fermi average gap $\Delta_{\Fs}$
is defined as the average value of the diagonal matrix element of the neutron
gap at the Fermi surface. The averaging procedure involves 10 levels above
$\mu_n$ and 10 levels below.  Furthermore, $\Delta_{\rm inf}$ means the
infinite neutron matter gap value found within the P2 model (the BCS value
$\Delta_n^{\rm BCS}(k_{\Fs})$ divided by 2) for the density $\rho$
corresponding to the Fermi momentum $k_{\rm F}^{\rm as}$. Finally
$\Delta^0_{\rm inf}$ is the same for the Fermi momentum $k_{\rm F}$. Let us
remind that the latter corresponds to the average nucleon density under
consideration. Obviously, the inequality $k_{\rm F}^{\rm as} < k_{\rm F}$ holds
because the WS cell contains a nuclear-like cluster in the center with the
density which exceeds the average one. Since in the small $k_{\rm F}$ region
the value of $\Delta_n^{\rm BCS}(k_{\Fs})$ increases with $k_{\rm F}$, the
inequality $\Delta_{\rm inf} < \Delta^0_{\rm inf}$ is valid. The difference is
especially large in the case of $k_{\rm F}{=}0.2\;$fm , which is nearby the
neutron drip point. Indeed, in this case almost all the matter is concentrated
in the central blob. The difference between the two neutron gaps decreases as
$k_{\rm F}$ increases.

 As it was discussed in \cite{P123}, the difference
between the asymptotic $\Delta_{\rm as}$ value and the infinite neutron matter
prediction $\Delta_{\rm inf}$ is a measure of validity of the LDA for the gap
calculation outside the central nuclear cluster. In the density region examined
in \cite{P123}, $k_{\rm F}{=}0.6 \div 1.2\;$fm$^{-1}$, the typical accuracy of
the LDA in this point was about 10\%. For $k_{\rm F}{=}0.4$ and $0.5
\;$fm$^{-1}$, it is of the same order, for $k_{\rm F}{=}0.2$ and $0.3
\;$fm$^{-1}$,  the difference is larger, which is an evidence of a stronger
proximity effect. The Fermi average value $\Delta_{\Fs}$ is usually very close
to $\Delta_{\rm as}$. This is because the region outside the nuclear cluster,
in which the function $\Delta(r)$ is almost a constant, gives the main
contribution to the matrix elements of $\Delta$ nearby the Fermi surface.

\begin{table}
\caption{ Characteristics of nuclear clusters for the equilibrium
configurations in the upper part of the inner crust.}
\bigskip

\begin{tabular}{|c|c|c|c|c|c|c|c|c|c|}
\hline
 $ k_{\Fs}$,fm$^{-1}$ & $Z$ & $A$ & $A_{cl}$ &  $x_{cl}$  & $y_{cl}$
& $R_n$,fm& $R_p$,fm & $R_m$,fm & $R_c$,fm \\
\hline
  0.2 & 52&  212  & 163 & 0.319 & 0.769 & 6.77 & 6.41 & 7.48  & 57.19 \\
  0.3 & 54&  562  & 174 & 0.310 & 0.309 & 6.88 & 6.55 & 7.65  & 52.79 \\
  0.4 & 50&  830  & 170 & 0.294 & 0.205 & 6.80 & 6.45 & 7.60  & 45.09 \\
  0.5 & 46& 1020  & 166 & 0.277 & 0.163 & 6.77 & 6.41 & 7.57  & 38.64 \\
\hline
\end{tabular}
\end{table}

\begin{table}
\caption{ Average characteristics (in MeV) of the neutron gap  in
the upper part of the inner crust.}
\bigskip

\begin{tabular}{|c|c|c|c|c|c|c|c|}
\hline
 $ k_{\Fs}$, fm$^{-1}$ & ~~$Z$~~  & $ k_{\Fs}^{\rm as}$, fm$^{-1}$ & $\Delta(0)$
  & $\Delta_{\rm as}$ & $\Delta_{\rm F}$ & $\Delta_{\rm inf}$ &
  $\Delta^0_{\rm inf}$ \\
\hline

  0.2 & 52 & 0.116   &  0.083   & 0.044  &  0.043 & 0.063 & 0.198  \\
  0.3 & 54 & 0.267   &  0.373   & 0.309  &  0.308 & 0.365 & 0.440 \\
  0.4 & 50 & 0.376   &  0.634   & 0.587  &  0.581 & 0.647 & 0.711 \\
  0.5 & 46 & 0.479   &  0.808   & 0.832  &  0.830 & 0.945 & 0.982 \\

\hline
\end{tabular}
\end{table}

\section{Conclusion}

Ground state properties of the upper part of the inner crust are
examined within the semi-microscopic self-consistent quantum
approach developed recently \cite{crust1,crust2,crust3,crust4}.
The theory is developed within the WS scheme and adopts a specific
effective density functional. The latter is constructed by
matching at the nuclear cluster surface the phenomenological
nuclear DF3 functional \cite{Fay} to the microscopic one
calculated for neutron matter within the Brueckner approach and
the Argonne force v$_{18}$. For the superfluid component of the
latter, the model P2 of \cite{P123} is used which takes into
account approximately the many-body corrections to the BCS
approximation. In this approach, the drip point separating the
inner crust from the outer one is determined mainly by the
phenomenological component of the semi-microscopic functional. For
the drip point region, we carried out two sets of calculations. In
the first one, the calculations are performed with only the
phenomenological DF3 functional, while in the second one the full
semi-microscopic of Eq.~(\ref{tot}) is employed.

The drip point density is found to be equal to $\rho_{\rm d}
\simeq 3.3 \cdot 10^{11}$g/cm$^3$ in the first case and $\rho_{\rm
d} \simeq 4 \cdot 10^{11}$g/cm$^3$ in the second. The first value
is a slightly smaller than the one which is commonly adopted
\cite{Rust}, while the second one practically coincides with it.
More significant difference appears for the  proton number $Z$ of
the nucleus at the drip point which is shifted from $Z{=}34\div
38$ of \cite{Rust} to $Z{=}52$. The corresponding neutron number
is about equal to the magic value $N{=}126$. This difference
originates, evidently, from the peculiarities of the effective DF3
force in comparison with, say, Skyrme forces, which are mainly
used in \cite{Rust}. For describing nuclei far from the
beta-stability valley which are considered in vicinity of the drip
point, the isovector component of the effective force is
especially important. The density dependence of the latter in
\cite{Fay} and that of the Skyrme forces is quite different. It
should be stressed that the parameters of \cite{Fay} were chosen
to reproduce the binding energies and the radii of long isotopic
chains, therefore one may hope that their extrapolation to the
drip line region is sufficiently accurate. The high accuracy of
the  effective force by S.A. Fayans et al., in comparison with the
popular version SLy4 of the Skyrme force, was recently confirmed
also by Y.Yu and A.Bulgac \cite{Bul1}.

In addition, we systematically examined the equilibrium configurations of the
upper part of the inner crust, for the density region corresponding to average
Fermi momenta $k_{\rm F}{=}0.2 \div 0.5\;$fm$^{-1}$. These calculations
supplement the ones of \cite{crust4,P123} to make the description of the ground
state of the inner crust of neutron star complete over all the density interval
except the bottom layers $k_{\rm F}> 1.2 \;$fm$^{-1}$ where the non-spherical
configurations are commonly believed \cite{Pet} to be energetically favored.
The features of the solution are in many ways similar to those for the $k_{\rm
F}{=}0.6 \div 1.2\;$fm$^{-1}$ examined in \cite{crust4,P123}, but there are
some peculiarities. In particular, in the density region under consideration,
for the nuclear-type clusters in the center of the WS cell there exist the
counterparts among ordinary atomic nuclei below the neutron drip
line.\footnote{The one for $k_{\rm F}{=}0.5\;$fm$^{-1}$, the $^{46}_{166}$Pd
nucleus, is situated almost exactly on the drip line.} This makes possible
direct comparison of the neutron and proton densities in the cluster with those
in the nuclear counterpart. It turned out that the corresponding density
distributions are quite similar, but  the values of the radius and  of the
diffuseness parameter for clusters are slightly bigger, the effect increasing
at larger $k_{\rm F}$.

As to the equilibrium  configurations ($Z,R_c$), they differ from the NV ones
significantly. Indeed, we found $Z{=}46\div 54$, whereas the value of $Z{=}40$
was found in \cite{NV} for all the region under consideration. To this respect,
the situation for higher values of $k_{\rm F}$ is similar \cite{crust4,P123},
but at $k_{\rm F}\simeq 1\;$fm$^{-1}$ we obtained $Z$ values which are
significantly smaller than the NV ones. As to the pairing gap function, it
shows a strong interplay between the region inside the cluster in the center of
the WS cell and the neutron matter outside it. This effect takes place for all
values of $k_{\rm F}$ including the smallest one where the gap value itself is
very small.

\vskip 0.5 cm

 This research was partially supported by the Grant
NSh-8756.2006.2 of the Russian Ministry for Science and Education
and by the RFBR grant 06-02-17171-a.

{}

\begin{figure}[h!]
\includegraphics [height=180mm,width=100mm]{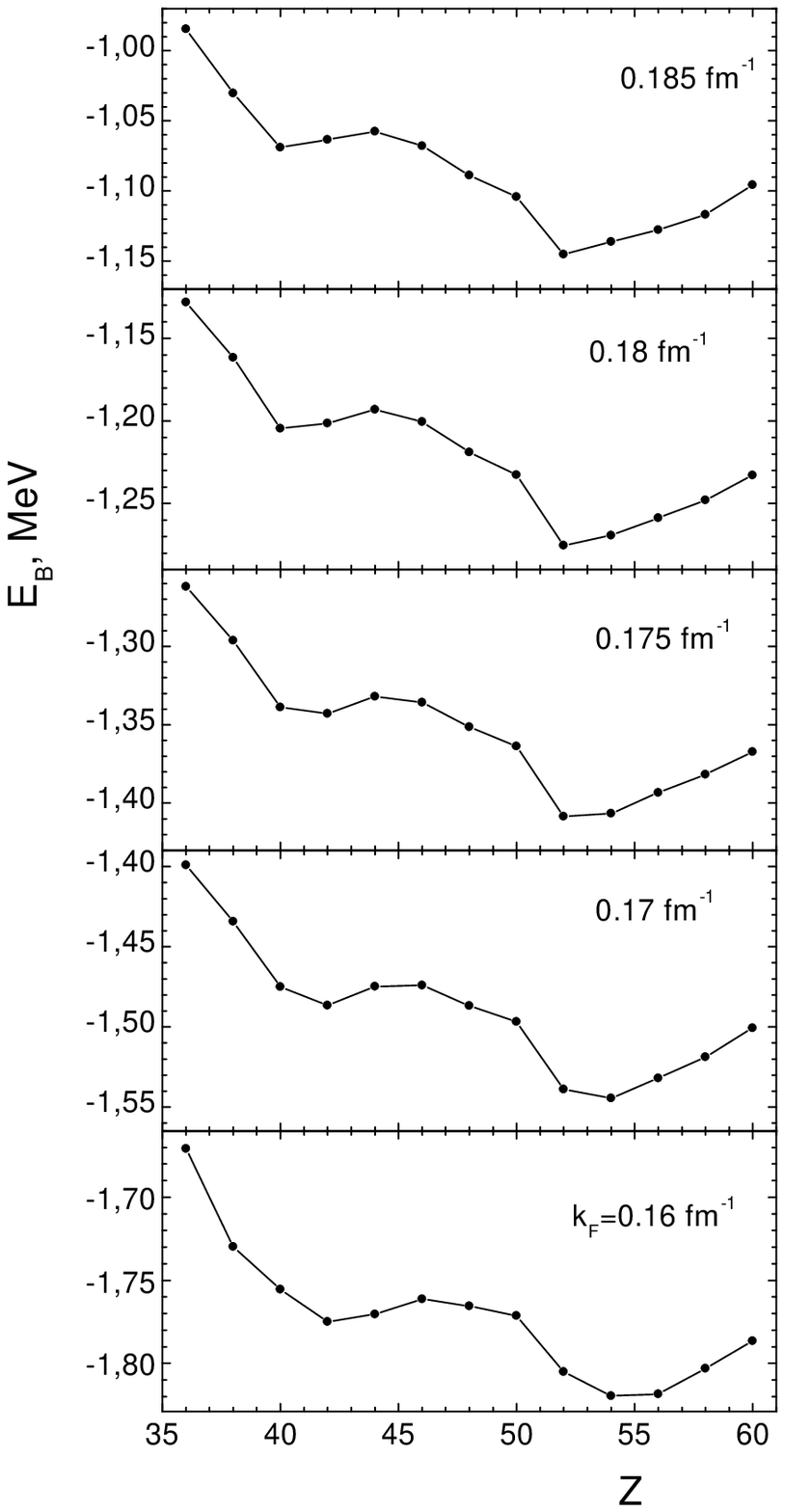}
\vspace{-2mm} \caption{Binding energy per a nucleon as a function
of $Z$ for various $k_{\rm F}$ values in vicinity of the drip
point.}
\end{figure}

\begin{figure}[h!]
\includegraphics [height=80mm,width=100mm]{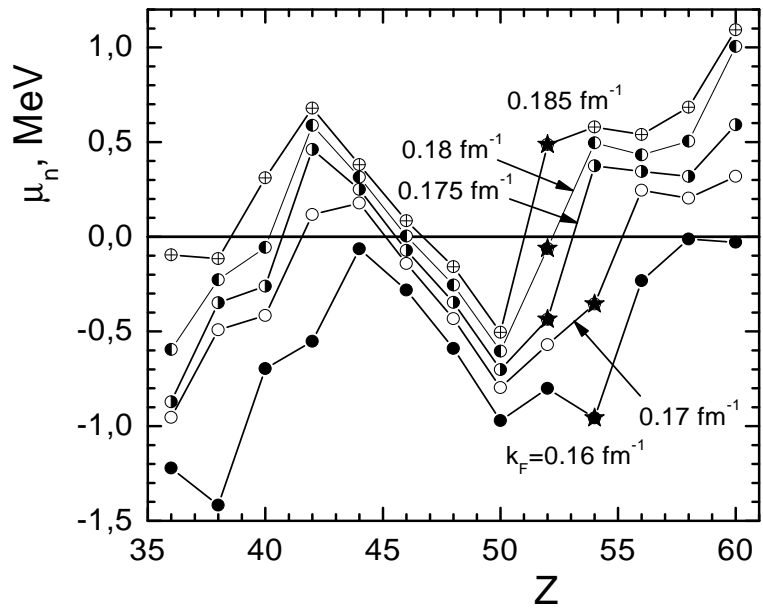}
\vspace{-2mm} \caption{ The neutron chemical potential as a
function of $Z$ for various $k_{\rm F}$ values in vicinity of the
drip point.}
\end{figure}

\begin{figure}[h!]
\includegraphics [height=120mm,width=100mm]{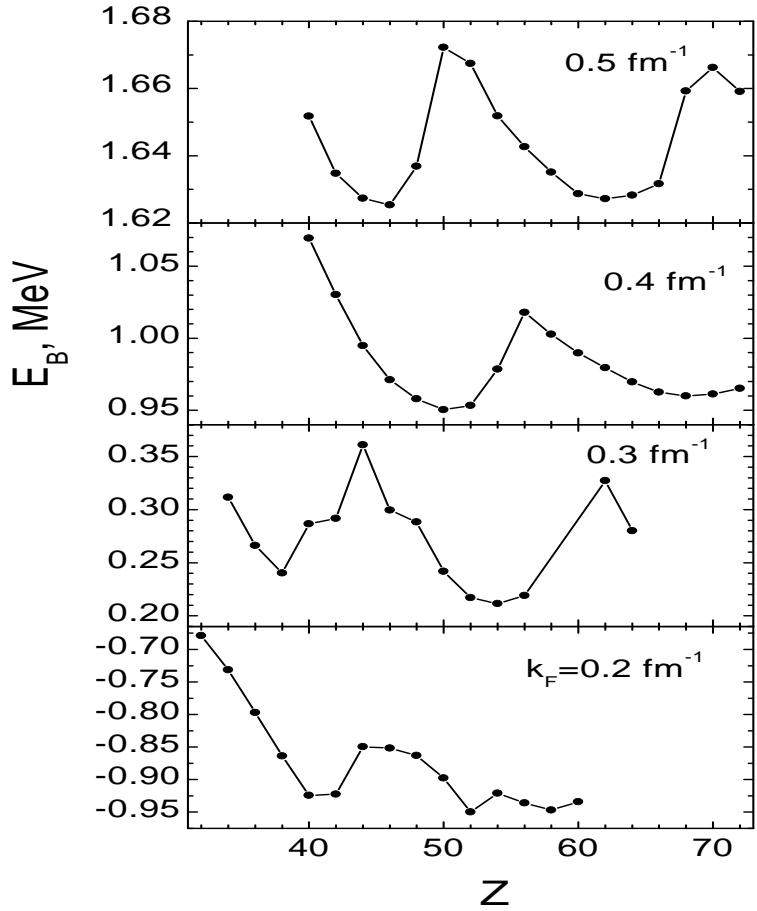}
\vspace{-2mm} \caption{Binding energy per a nucleon as a function
of $Z$ for various $k_{\rm F}$ values in the upper part of the
inner crust.}
\end{figure}

\begin{figure}[h!]
\includegraphics [height=100mm,width=150mm]{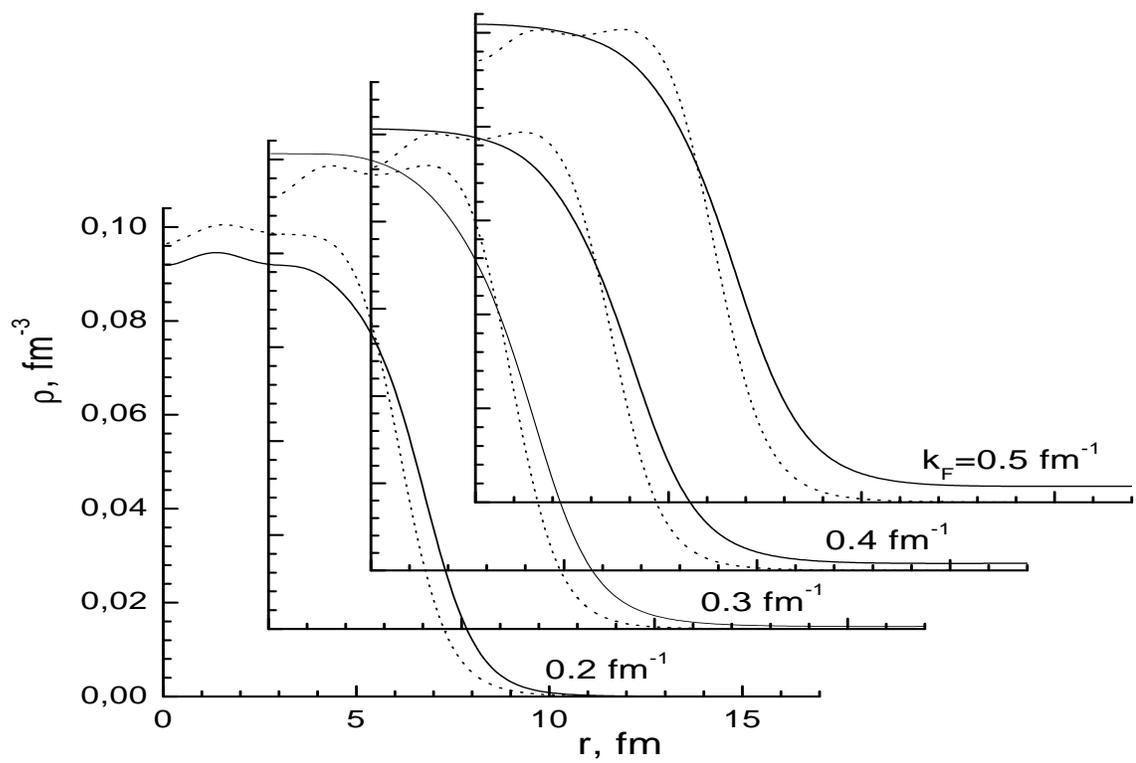}
\vspace{-2mm} \caption{ The neutron density distributions  for
various $k_{\rm F}$ values.}
\end{figure}

\begin{figure}[h!]
\includegraphics [height=100mm,width=100mm]{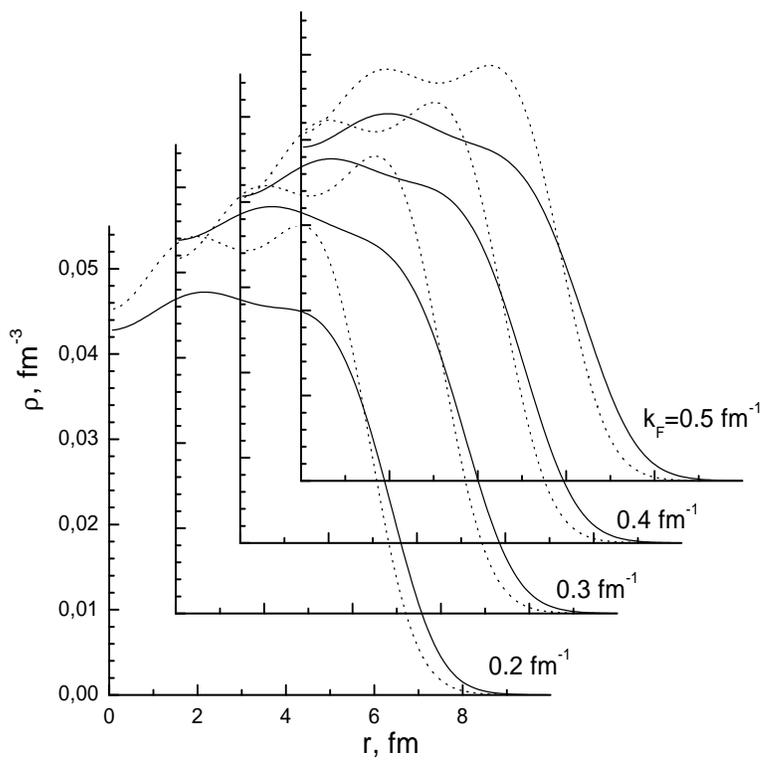}
\vspace{-2mm} \caption{ The proton density distributions  for
various $k_{\rm F}$ values.}
\end{figure}

\begin{figure}[h!]
\includegraphics [height=100mm,width=100mm]{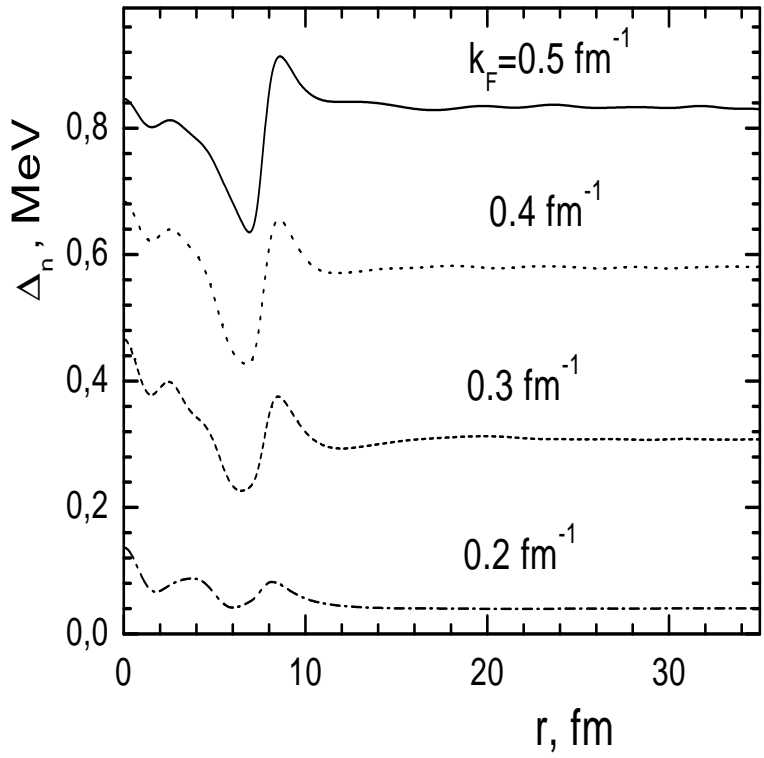}
\vspace{-2mm} \caption{ The neutron gap functions  for various
$k_{\rm F}$ values.}
\end{figure}

\end{document}